\documentclass[12pt]{article}
\usepackage{amssym}

\newcommand{\etap}{\eta^{\dagger}}
\newcommand{\half}{\frac{1}{2}}
\newcommand{\rr}{\mbox{$\Bbb R$}}
\newcommand{\htilde}{\tilde{h}}
\newcommand{\hht}{\tilde{H}}
\newcommand{\rhot}{\tilde{\rho}}
\newcommand{\omegat}{\tilde{\omega}}
\def\sech{\mathop{\rm sech}\nolimits}

\title{
Pseudo-Hermiticity and some consequences of a generalized quantum condition}
\author{B Bagchi$^a$, C Quesne$^{b}$, R Roychoudhury$^c$\\ 
{\small
$^a$ Department of Applied Mathematics, University of Calcutta,} \\ {\small 92 Acharya
Prafulla Chandra Road, Kolkata 700 009, India}\\ 
{\small $^b$ Physique Nucl\'eaire
Th\'eorique et Physique Math\'ematique,  Universit\'e Libre de Bruxelles,} \\ 
{\small Campus de la Plaine CP229, Boulevard~du Triomphe, B-1050
Brussels, Belgium}\\
{\small $^c$ Physics and Applied Mathematics Unit, Indian Statistical Institute, Kolkata
700 035, India}\\ {\small E-mail: bbagchi123@rediffmail.com, cquesne@ulb.ac.be,
raj@isical.ac.in}}
\date{ }
\begin{document}
\baselineskip=22pt plus 1pt minus 1pt
%%%%%%%%%%%%%%%%%%%%%%%%%%%%%%%%%%%%%%%%%%%%%%%%%%%%%%%%%%
\maketitle

\begin{abstract} 
We exploit the hidden symmetry structure of a recently proposed non-Hermitian
Hamiltonian and of its Hermitian equivalent one. This sheds new light on the
pseudo-Hermitian character of the former and allows access to a generalized quantum
condition. Special cases lead to hyperbolic and Morse-like potentials in the framework of
a coordinate-dependent mass model. 
\end{abstract}


%
%========================================================================
%
\newpage
Non-Hermitian Hamiltonians are currently an active field of research~\cite{bender98,
delabaere, bender99, levai, bagchi00, znojil, khare, ahmed, dorey, mosta02, solombrino,
bender03, bender04, trinh, sree}, motivated by the necessity to understand the
mathematical properties of their subclass, namely the pseudo-Hermitian or the
$PT$-symmetric ones. Also, to investigate the existence of a suitable similarity
transformation that maps such Hamiltonians to an equivalent Hermitian form is important
from a physical point of view. A consistent theory of quantum mechanics demands a
certain inner product that ensures the associated norm to be positive definite. In this
direction there have been efforts to look for non-Hermitian Hamiltonians which have a
real spectrum such that the accompanying dynamics is unitary~\cite{bender04}.\par
%
%-----------------------------------------------------------------------------------
%
It must be said that the interest in non-Hermitian Hamiltonians was stepped up by a
conjecture of Bender and Boettcher~\cite{bender98} that $PT$-symmetric Hamiltonians
could possess real bound-state eigenvalues. Subsequently, Mostafazadeh~\cite{mosta02}
showed that the concept of $PT$ symmetry has its roots in the theory of
pseudo-Hermitian operators. He pointed out that the reality of the spectrum is
ensured~\cite{mosta04} if the Hamiltonian $H$ is Hermitian with respect to a
positive-definite inner product
$\langle\cdot, \cdot\rangle_+$ on the Hilbert space $\cal H$ in which $H$ is acting.
This inner product may be expressed in terms of the defining inner product $\langle\cdot,
\cdot\rangle$ as
\begin{equation}
  \langle\cdot, \cdot\rangle_+ = \langle\cdot, \zeta_+ \cdot\rangle  \label{eq:inner}
\end{equation}
where the metric operator $\zeta_+: {\cal H} \to {\cal H}$ belongs to the set of all 
Hermitian invertible operators and, in itself, is positive definite. The pseudo-Hermiticity of
the Hamiltonian $H$ given by
\begin{equation}
  H^{\dagger} = \zeta_+ H \zeta_+^{-1}
\end{equation}
serves as one of the plausible necessary and sufficient conditions for the reality of the
spectrum.\par
%
%------------------------------------------------------------------------------------
%
The Hilbert space equipped with the inner product (\ref{eq:inner}) is identified as the
physical Hilbert space ${\cal H}_{\rm phys}$. Any observable $O \in {\cal H}_{\rm
phys}$ is related to the Hermitian operator $o \in \cal H$ by means of a similarity
transformation
\begin{equation}
  O = \rho^{-1} o \rho
\end{equation}
where $\rho = \sqrt{\zeta_+}$ is unitary and maps ${\cal H}_{\rm phys} \to \cal H$.
For the Hermitian Hamiltonian $h$ we can write
\begin{equation}
  h = \rho H \rho^{-1}  \label{eq:h}
\end{equation}
which speaks of the quasi-Hermiticity of $H$~\cite{scholtz, kretschmer} and $h$ may be
looked upon as the equivalent Hermitian analogue of $H$.\par
%
%-----------------------------------------------------------------------------
%  
Some studies have been done~\cite{mosta04, jones, mosta05, ghosh, banerjee} to
arrive at a specific form for the mapping function $\rho$ for various non-Hermitian
quantum systems. Very recently, in an enlightening paper, Jones~\cite{jones} considered
different possibilities for $\rho = \rho(x)$ and demonstrated, in particular, that a
non-Hermitian Hamiltonian, which was earlier proposed and solved by
Swanson~\cite{swanson} using operator techniques, admits of an equivalent
Hermitian representation by applying a similarity transformation. Swanson's Hamiltonian
reads
\begin{equation}
  H^{(\alpha, \beta)} = \Bigl(H^{(\beta, \alpha)}\Bigr)^{\dagger} = \omega \etap \eta
  + \alpha \eta^2 + \beta \eta^{\dagger2} + \half \omega \qquad \omega,
  \alpha, \beta \in \rr  \label{eq:swanson}
\end{equation}
with $\eta$ and $\etap$ obeying the standard commutation relation $[\eta, \etap] = 1$.
It is obviously Hermitian only if $\alpha = \beta$, but $PT$-symmetric (or, equivalently,
$P$-pseudo-Hermitian) for all values of $\alpha$ and $\beta$. Jones' transformation of
$H^{(\alpha, \beta)}$ to its equivalent Hermitian analogue is reminiscent of partial
algebraization method to tackle quasi-solvable systems through an imaginary gauge
transformation (for a review see~\cite{shifman}). Indeed, $H^{(\alpha, \beta)}$ can be
presented as a linear combination of sl(2) generators $K_0 = \half(\etap \eta + \half)$,
$K_+ = \half \eta^{\dagger2}$, $K_- = \half \eta^2$.\par
%
%----------------------------------------------------------------------------------
%
An important point we wish to observe here is that the Hermitian Hamiltonian $h^{(\alpha,
\beta)}$ corresponding to Swanson's Hamiltonian whenever $\omega > \alpha + \beta$,
namely~\cite{jones}
\begin{eqnarray}
  h^{(\alpha,\beta)} & = & - \half (\omega - \alpha - \beta) \frac{d^2}{dx^2} + \half
       \frac{\omega^2 - 4\alpha\beta}{\omega - \alpha - \beta} x^2 \nonumber \\
  & = & \rho_{(\alpha,\beta)} H^{(\alpha, \beta)} \rho_{(\alpha,\beta)}^{-1}
       \label{eq:h-swanson}
\end{eqnarray} 
where 
\begin{equation}
  \rho_{(\alpha,\beta)} = \exp\left(- \half \frac{\alpha-\beta}{\omega-\alpha-\beta}
  x^2\right)  \label{eq:rho-swanson}
\end{equation}
is clearly symmetric in the parameters $\alpha$ and $\beta$, i.e.,
$h^{(\alpha,\beta)} = h^{(\beta,\alpha)}$, a feature not present in (\ref{eq:swanson}).
A consequence of this is that
\begin{equation}
  h^{(\alpha,\beta)} = \rho_{(\beta,\alpha)} H^{(\beta,\alpha)}
  \rho_{(\beta,\alpha)}^{-1} = \rho_{(\beta,\alpha)} H^{(\alpha,\beta)\dagger}
  \rho_{(\beta,\alpha)}^{-1}  \label{eq:property}  
\end{equation}
whence it follows from (\ref{eq:h-swanson}) that
\begin{equation}
  \left[\rho_{(\beta,\alpha)}^{-1} \rho_{(\alpha,\beta)}\right] H^{(\alpha, \beta)}
  \left[\rho_{(\alpha,\beta)}^{-1} \rho_{(\beta,\alpha)}\right] =
  H^{(\alpha,\beta)\dagger}. 
\end{equation}
The above equation clearly shows $H^{(\alpha,\beta)}$ to be $\zeta$-pseudo-Hermitian
if we define $\zeta = \rho_{(\beta,\alpha)}^{-1} \rho_{(\alpha,\beta)}$. A sufficient
condition for the positivity of $\zeta$ may be provided by the restriction
$\rho_{(\alpha,\beta)} = \rho_{(\beta,\alpha)}^{-1}$, which is obeyed by
(\ref{eq:rho-swanson}).\par
%
%------------------------------------------------------------------------------------
%    
The purpose of this Letter is to study the Hamiltonian (\ref{eq:swanson}) from the point
of view of a generalized quantum condition for which $[\eta, \etap] \ne 1$. Although
$PT$ symmetry is generally lost in this way, we will show that the transformed
Hamiltonian still remains pseudo-Hermitian with respect to a positive-definite $\zeta_+$
and a corresponding Hermitian counterpart can be easily set up. On the other hand, a
generalized quantum condition allows access to those physical systems that are
underlined by a coordinate dependence in mass by suitably choosing $\eta$.\par
%
%-----------------------------------------------------------------------------
% 
Let us adopt the most general first-order differential form for $\eta$, namely
\begin{equation}
  \eta = a(x) \frac{d}{dx} + b(x) \qquad a(x), b(x) \in \rr.  \label{eq:eta}
\end{equation}
It yields for the commutator
\begin{equation}
  [\eta, \etap] = 2ab' - aa''  \label{eq:commut}
\end{equation}
where a prime denotes derivative with respect to $x$. Then the corresponding eigenvalue
equation for (\ref{eq:swanson}) reads  
\begin{equation}
  \hht^{(\alpha,\beta)} \phi(x) \equiv \left[- \omegat \frac{d}{dx} a^2 \frac{d}{dx} +
  (\omegat aa' + c_1) \frac{d}{dx} + c_2\right] \phi(x) = E \phi(x)  \label{eq:hht} 
\end{equation}
where the functions $c_1(x)$ and $c_2(x)$ are given by
\begin{eqnarray}
  c_1(x) & = & - \omegat aa' + (\alpha - \beta) a (2b-a') \label{eq:c1} \\
  c_2(x) & = & \omegat (b^2 - ab' - a'b) + \alpha b(2b - a') + \beta [(b-a')(2b - a') -
      a(2b'-a'')] \nonumber \\
  && \mbox{} + \half (\omegat + \alpha + \beta)  \label{eq:c2} 
\end{eqnarray}
along with $\omegat = \omega - \alpha - \beta$. To be able to bring equation
(\ref{eq:hht}) to a Schr\"odinger form, it is appropriate to assume $\omegat > 0$ or,
equivalently, $\omega > \alpha + \beta$. Without loss of generality, we may set
$\omegat = 1$. Notice that with the form (\ref{eq:eta}), $\hht^{(\alpha,\beta)}$ is not
$PT$-symmetric unless $a(x)$ is an odd function and $b(x)$ is an even function of
$x$.\par
%
%-----------------------------------------------------------------------------------
% 
It is straightforward to obtain the Hermitian analogue of (\ref{eq:hht}) by removing the
first-derivative term with the help of a similarity transformation. This necessitates
defining $\phi(x) = w(x) \chi(x)$, $w(x) \equiv \rhot_{(\alpha,\beta)}^{-1}$, and
imposing the constraint 
\begin{equation}
  \rhot_{(\alpha,\beta)} = a^{-1/2} \exp\left(- \half \int^x \frac{c_1}{a^2}\, dx'\right)
\end{equation}
provided this function is well defined on $\rr$. We then obtain a Hermitian equivalent
form for $\hht^{(\alpha,\beta)}$ in a manner similar to (\ref{eq:h}):
\begin{equation}
  \htilde^{(\alpha,\beta)} = \rhot_{(\alpha,\beta)} \hht^{(\alpha,\beta)}
  \rhot_{(\alpha,\beta)}^{-1}
\end{equation}
where $\htilde^{(\alpha,\beta)}$ reads explicitly
\begin{eqnarray}
  \htilde^{(\alpha,\beta)} & = & - \frac{d}{dx} a^2 \frac{d}{dx} + V_{\rm eff}(x)
        \label{eq:htilde} \\
  V_{\rm eff}(x) & = & - a^2 \frac{w''}{w} -
       (aa' - c_1) \frac{w'}{w} + c_2. 
\end{eqnarray}
\par
%
%---------------------------------------------------------------------------------------
% 
Substituting for $c_1$ and $c_2$ by their expressions in (\ref{eq:c1}) and
(\ref{eq:c2}), $\rhot_{(\alpha,\beta)}$ and $V_{\rm eff}$ can be expressed as
\begin{eqnarray}
  \rhot_{(\alpha,\beta)} & = & a^{\half (\alpha-\beta)} \exp\left(- (\alpha-\beta)
       \int^x \frac{b}{a}\, dx'\right)  \label{eq:rhotilde} \\
  V_{\rm eff} & = & \half (\alpha+\beta) aa'' + \left[\half (\alpha+\beta) + \frac{1}{4}
       (\alpha-\beta)^2\right] a^{\prime2} - \left[1 + 2(\alpha+\beta) +  (\alpha-\beta)^2
       \right] a'b \nonumber \\
  && \mbox{} + \left[1 + 2(\alpha+\beta) +  (\alpha-\beta)^2 \right] b^2 - (\alpha+
       \beta+1) ab' + \half (\alpha+\beta+1).  \label{eq:Veff}  
\end{eqnarray}
Note that $\rhot_{(\alpha,\beta)} = 1$ for $\alpha = \beta$, which is as it should be.
Specifically, for $\alpha = \beta = 0$, $V_{\rm eff}$ reduces to the form
\begin{equation}
  V_{\rm eff} = b^2 - (ab)' + \half 
\end{equation}
which has been considered before in the context of a coordinate-dependent mass
Schr\"odinger equation~\cite{bagchi04}.\par
%
%-----------------------------------------------------------------------------
%
We now proceed to show that for a positive function $a(x)$, $\hht^{(\alpha,\beta)}$ is
pseudo-Hermitian with respect to some positive-definite $\zeta_+$ and has therefore a
real spectrum. For this we need to notice that $\htilde^{(\alpha,\beta)}$ in
(\ref{eq:htilde}) is symmetric with respect to the parameters $\alpha$ and $\beta$,
given the form for $V_{\rm eff}$ in (\ref{eq:Veff}). As such, equation
(\ref{eq:property}) holds in this case too with $H^{(\alpha,\beta)}$ and
$\rho_{(\alpha,\beta)}$ replaced by $\hht^{(\alpha,\beta)}$ and
$\rhot_{(\alpha,\beta)}$, respectively. Moreover, from (\ref{eq:rhotilde}), the condition
$\rhot_{(\alpha,\beta)} = \rhot_{(\beta,\alpha)}^{-1}$ also holds, so that
$\hht^{(\alpha,\beta)}$ is pseudo-Hermitian,
\begin{equation}
  \hht^{(\alpha,\beta)\dagger} = \zeta_+ \hht^{(\alpha,\beta)} \zeta_+^{-1}
\end{equation}
with $\zeta_+ = a^{\alpha-\beta} \exp\left[- 2(\alpha-\beta) \int^x (b/a)\, dx'\right] >
0$ if $a(x)>0$. For example, for coordinate-dependent mass systems, $a(x)$ could be
identified with the inverse square root of a certain mass function that is strictly positive
definite. Indeed, a time-independent Schr\"odinger equation in the presence of a
coordinate-dependent mass can always be brought~\cite{bagchi04} to the Hermitian form
\begin{equation}
  \left(- \frac{d}{dx} \frac{1}{M(x)} \frac{d}{dx} + V_{\rm eff}(x)\right) \psi(x) = E
  \psi(x)
\end{equation}
where we have set $m(x) = m_0 M(x)$ to make $M(x)$ dimensionless. We also used units
such that $\hbar = 2m_0 = 1$. A comparison with (\ref{eq:htilde}) at once reveals that
$a(x) = [M(x)]^{-1/2}$ as we just noted.\par
%
%------------------------------------------------------------------------------------
%
Let us assume, for instance, $a(x) = \cosh qx$ (corresponding to the solitonic profile
$M(x) = \sech^2 qx$) and $b(x) = \kappa q \sinh qx$ with $q>0$ and $\kappa > \half$.
Then the generalized quantum condition reads $[\eta, \etap] = (2\kappa-1) q^2
\cosh^2 qx$ and we get from (\ref{eq:rhotilde}) and (\ref{eq:Veff})
\begin{equation}
  \rhot_{(\alpha,\beta)} = (\cosh qx)^{-(\alpha-\beta) (\kappa-\half)} \qquad
  V_{\rm eff} = \frac{1}{4} q^2 (2\lambda+1) (2\lambda-3) \cosh^2 qx + V_0
\end{equation}
where $V_0$ is some constant, $\lambda \equiv \half + \sqrt{\Delta}$ and $\Delta
\equiv (\kappa-1)^2 + (\kappa-1) (2\kappa-1) (\alpha+\beta) + (\kappa-\half)^2
(\alpha-\beta)^2$. For those values of $\kappa$, $\alpha$, $\beta$ for which $\Delta >
0$, and therefore $\lambda$ is real and positive, the resulting Hermitian Hamiltonian
$\htilde^{(\alpha,\beta)}$ of equation (\ref{eq:htilde}) has an infinite number of bound
states, corresponding to the energy eigenvalues $E_n = q^2 (n+\lambda-\half)
(n+\lambda+\half) + V_0$ and the wavefunctions $\chi_n(x) \propto (\sech
qx)^{\lambda+\half} C_n^{(\lambda)}(\tanh qx)$ (with $C_n^{(\lambda)}$ denoting a
Gegenbauer polynomial)~\cite{bagchi05}. We conclude that the associated
non-Hermitian Hamiltonian $\hht^{(\alpha,\beta)}$ has the same real spectrum with
wavefunctions $\phi_n(x) = \rhot_{(\alpha,\beta)}^{-1} \chi_n(x)$.\par
%
%-------------------------------------------------------------------------------
%
{}Furthermore, it is interesting to consider the canonical condition $[\eta, \etap] = 1$
for the general choice (\ref{eq:eta}). From (\ref{eq:commut}), we get for $b$ the
solution
\begin{equation}
  b = - \frac{g''}{2g^{\prime2}} + \half g + \mu  \label{eq:b}
\end{equation}
where we have defined $g(x) = \int^x dx' /a(x')$ and $\mu$ is an integration
constant. For example, for the same preceding choice of $M(x)$, the function $b$ turns
out to be
\begin{equation}
  b = \half q \sinh qx + \frac{1}{q} \tan^{-1}[\exp(qx)]
\end{equation}
up to an additive integration constant.\par
%
%-------------------------------------------------------------------------------------------
%
In the context of the effective potential (\ref{eq:Veff}), we obtain on using (\ref{eq:b})
\begin{equation}
  V_{\rm eff} = \half \frac{g'''}{g^{\prime3}} - \frac{5}{4}
  \frac{g^{\prime\prime2}}{g^{\prime4}} + [1 + 2(\alpha+\beta) + (\alpha-\beta)^2]
  \left(\half g + \mu\right)^2.
\end{equation}
\par
%
%----------------------------------------------------------------------------
%
On choosing $g$ we can get explicit forms for $\rhot_{(\alpha,\beta)}$ and
$V_{\rm eff}$. For instance, if $g = - \frac{1}{p} e^{-px}$, $p \in \rr$, then
$\rhot_{(\alpha,\beta)}$ is given by
\begin{equation}
  \rhot_{(\alpha,\beta)} = \exp\left[- (\alpha-\beta) \left(- \frac{1}{2p} e^{-px} + 
  \mu\right)^2\right]
\end{equation}
and $V_{\rm eff}$ describes a Morse-like potential
\begin{equation}
  V_{\rm eff } = - \frac{3}{4} p^2 e^{2px} + [1 + 2(\alpha+\beta) + (\alpha-\beta)^2]
  \left(- \frac{1}{2p} e^{-px} + \mu\right)^2.
\end{equation}
This is another example of a Hermitian equivalence derived by means of a similarity
transformation. Note that the mapping function $\rhot_{(\alpha,\beta)}$ is consistent
with an exponential mass background: $M = e^{-2px}$.\par
%
%---------------------------------------------------------------------------
% 
{}Finally, we can try to recast the general $V_{\rm eff}$ in the form
\begin{equation}
  V_{\rm eff} = b_1^2 - (ab_1)' + \xi  \label{eq:Veff-bis}
\end{equation}
where $b_1 = d_1 b + d_2 a'$ and $\xi$, $d_1$, $d_2$ are some real constants.
Comparing with (\ref{eq:Veff}) and $d_1$, $d_2$ in favour of $\alpha$, $\beta$, we
have two solutions, either
\begin{equation}
  b_1 = (1+\alpha) b - \half \alpha a' \qquad \xi = \half (1+\alpha) \qquad \beta = 0
  \label{eq:sol1}
\end{equation}
or
\begin{equation}
  b_1 = (1+\beta) b - \half \beta a' \qquad \xi = \half (1+\beta) \qquad \alpha = 0.
  \label{eq:sol2}
\end{equation}
Note that the symmetric nature of $V_{\rm eff}$ with respect to $\alpha$, $\beta$
interchange makes it evident to derive (\ref{eq:sol2}) from (\ref{eq:sol1}) and vice
versa. We can also infer from (\ref{eq:htilde}) and (\ref{eq:Veff-bis}) that in these
special cases there exists~\cite{bagchi04} an intertwining operator for
$\htilde^{(\alpha,\beta)}$ induced by $\eta_1 = a \frac{d}{dx} + b_1$.\par
%
%=========================================================
%
\section*{Acknowledgments}

Two of us, BB and RR, gratefully acknowledge the support of the National Fund for
Scientific Research (FNRS), Belgium, and the warm hospitality at PNTPM, Universit\'e Libre
de Bruxelles, where this work was carried out. CQ is a Research Director of the National
Fund for Scientific Research (FNRS), Belgium.\par
%
%==============================================
%

\end{document}